\documentclass[prl,twocolumn,superscriptaddress,showpacs,amssymb,amsmath,amsfonts,aps]{revtex4}
\usepackage{epsfig}
\usepackage{natbib}

\begin{document}
%
%
%
\title{{\large Upper limits for the photoproduction cross section for the $\Phi^{--}(1860)$ pentaquark state off the deuteron}}
%
%
%

\newcommand*{\UNH}{University of New Hampshire, Durham, New Hampshire 03824-3568}
\affiliation{\UNH}

\newcommand*{\JLAB}{Thomas Jefferson National Accelerator Facility, Newport News, Virginia 23606}
\affiliation{\JLAB}

\newcommand*{\SCAROLINA}{University of South Carolina, Columbia, South Carolina 29208}
\affiliation{\SCAROLINA}

\newcommand*{\RICE }{ Rice University, Houston, Texas 77005-1892} 
\affiliation{\RICE } 

\newcommand*{\ANL}{Argonne National Laboratory, Argonne, Illinois 60441}
\newcommand*{\ANLindex}{1}
\affiliation{\ANL}
\newcommand*{\ASU}{Arizona State University, Tempe, Arizona 85287-1504}
\newcommand*{\ASUindex}{2}
\affiliation{\ASU}
\newcommand*{\UCLA}{University of California at Los Angeles, Los Angeles, California  90095-1547}
\newcommand*{\UCLAindex}{3}
\affiliation{\UCLA}
\newcommand*{\CSUDH}{California State University, Dominguez Hills, Carson, CA 90747}
\newcommand*{\CSUDHindex}{4}
\affiliation{\CSUDH}
\newcommand*{\CANISIUS}{Canisius College, Buffalo, NY}
\newcommand*{\CANISIUSindex}{5}
\affiliation{\CANISIUS}
\newcommand*{\CMU}{Carnegie Mellon University, Pittsburgh, Pennsylvania 15213}
\newcommand*{\CMUindex}{6}
\affiliation{\CMU}
\newcommand*{\CUA}{Catholic University of America, Washington, D.C. 20064}
\newcommand*{\CUAindex}{7}
\affiliation{\CUA}
\newcommand*{\SACLAY}{CEA, Centre de Saclay, Irfu/Service de Physique Nucl\'eaire, 91191 Gif-sur-Yvette, France}
\newcommand*{\SACLAYindex}{8}
\affiliation{\SACLAY}
\newcommand*{\CNU}{Christopher Newport University, Newport News, Virginia 23606}
\newcommand*{\CNUindex}{9}
\affiliation{\CNU}
\newcommand*{\UCONN}{University of Connecticut, Storrs, Connecticut 06269}
\newcommand*{\UCONNindex}{10}
\affiliation{\UCONN}
\newcommand*{\EDINBURGH}{Edinburgh University, Edinburgh EH9 3JZ, United Kingdom}
\newcommand*{\EDINBURGHindex}{11}
\affiliation{\EDINBURGH}
\newcommand*{\FU}{Fairfield University, Fairfield CT 06824}
\newcommand*{\FUindex}{12}
\affiliation{\FU}
\newcommand*{\FIU}{Florida International University, Miami, Florida 33199}
\newcommand*{\FIUindex}{13}
\affiliation{\FIU}
\newcommand*{\FSU}{Florida State University, Tallahassee, Florida 32306}
\newcommand*{\FSUindex}{14}
\affiliation{\FSU}
\newcommand*{\Genova}{Universit$\grave{a}$ di Genova, 16146 Genova, Italy}
\newcommand*{\Genovaindex}{15}
\affiliation{\Genova}
\newcommand*{\GWUI}{The George Washington University, Washington, DC 20052}
\newcommand*{\GWUIindex}{16}
\affiliation{\GWUI}
\newcommand*{\ISU}{Idaho State University, Pocatello, Idaho 83209}
\newcommand*{\ISUindex}{17}
\affiliation{\ISU}
\newcommand*{\INFNFE}{INFN, Sezione di Ferrara, 44100 Ferrara, Italy}
\newcommand*{\INFNFEindex}{18}
\affiliation{\INFNFE}
\newcommand*{\INFNFR}{INFN, Laboratori Nazionali di Frascati, 00044 Frascati, Italy}
\newcommand*{\INFNFRindex}{19}
\affiliation{\INFNFR}
\newcommand*{\INFNGE}{INFN, Sezione di Genova, 16146 Genova, Italy}
\newcommand*{\INFNGEindex}{20}
\affiliation{\INFNGE}
\newcommand*{\INFNRO}{INFN, Sezione di Roma Tor Vergata, 00133 Rome, Italy}
\newcommand*{\INFNROindex}{21}
\affiliation{\INFNRO}
\newcommand*{\ORSAY}{Institut de Physique Nucl\'eaire ORSAY, Orsay, France}
\newcommand*{\ORSAYindex}{22}
\affiliation{\ORSAY}
\newcommand*{\ITEP}{Institute of Theoretical and Experimental Physics, Moscow, 117259, Russia}
\newcommand*{\ITEPindex}{23}
\affiliation{\ITEP}
\newcommand*{\JMU}{James Madison University, Harrisonburg, Virginia 22807}
\newcommand*{\JMUindex}{24}
\affiliation{\JMU}
\newcommand*{\KNU}{Kyungpook National University, Daegu 702-701, Republic of Korea}
\newcommand*{\KNUindex}{25}
\affiliation{\KNU}
\newcommand*{\LPSC}{LPSC, Universite Joseph Fourier, CNRS/IN2P3, INPG, Grenoble, France}
\newcommand*{\LPSCindex}{26}
\affiliation{\LPSC}

\newcommand*{\NSU}{Norfolk State University, Norfolk, Virginia 23504}
\newcommand*{\NSUindex}{28}
\affiliation{\NSU}

\newcommand*{\OHIOU}{Ohio University, Athens, Ohio  45701}
\newcommand*{\OHIOUindex}{28}
\affiliation{\OHIOU}
\newcommand*{\ODU}{Old Dominion University, Norfolk, Virginia 23529}
\newcommand*{\ODUindex}{29}
\affiliation{\ODU}

\newcommand*{\RPI}{Rensselaer Polytechnic Institute, Troy, New York 12180-3590}
\newcommand*{\RPIindex}{30}
\affiliation{\RPI}
\newcommand*{\URICH}{University of Richmond, Richmond, Virginia 23173}
\newcommand*{\URICHindex}{31}
\affiliation{\URICH}
\newcommand*{\ROMAII}{Universita' di Roma Tor Vergata, 00133 Rome Italy}
\newcommand*{\ROMAIIindex}{32}
\affiliation{\ROMAII}
\newcommand*{\MSU}{Skobeltsyn Nuclear Physics Institute, Skobeltsyn Nuclear Physics Institute, 119899 Moscow, Russia}
\newcommand*{\MSUindex}{33}
\affiliation{\MSU}

\newcommand*{\UTFSM}{Universidad T\'{e}cnica Federico Santa Mar\'{i}a, Casilla 110-V Valpara\'{i}so, Chile}
\newcommand*{\UTFSMindex}{36}
\affiliation{\UTFSM}
\newcommand*{\GLASGOW}{University of Glasgow, Glasgow G12 8QQ, United Kingdom}
\newcommand*{\GLASGOWindex}{37}
\affiliation{\GLASGOW}
\newcommand*{\VIRGINIA}{University of Virginia, Charlottesville, Virginia 22901}
\newcommand*{\VIRGINIAindex}{38}
\affiliation{\VIRGINIA}
\newcommand*{\WM}{College of William and Mary, Williamsburg, Virginia 23187-8795}
\newcommand*{\WMindex}{39}
\affiliation{\WM}
\newcommand*{\YEREVAN}{Yerevan Physics Institute, 375036 Yerevan, Armenia}
\newcommand*{\YEREVANindex}{40}
\affiliation{\YEREVAN}

\newcommand*{\NOWANL}{Argonne National Laboratory, Argonne, Illinois 60441}
\newcommand*{\NOWLANL}{Los Alamos National Laborotory, New Mexico, NM}
\newcommand*{\NOWMSU}{Skobeltsyn Nuclear Physics Institute, Skobeltsyn Nuclear Physics Institute, 119899 Moscow, Russia}
\newcommand*{\NOWINFNGE}{INFN, Sezione di Genova, 16146 Genova, Italy}

\newcommand*{\DECEASED}{Deceased}


%
%
\author{H.~Egiyan}
     \affiliation{\UNH}
     \affiliation{\JLAB}
\author{J.~Langheinrich}
     \affiliation{\SCAROLINA}
\author {R.W.~Gothe} 
     \affiliation{\SCAROLINA}
\author{L. Graham}
     \affiliation{\SCAROLINA}
\author{M.~Holtrop}
     \affiliation{\UNH}
\author{H.~Lu}
     \affiliation{\CMU}
     \affiliation{\SCAROLINA}
\author{P.~Mattione}
     \affiliation{\CMU}
     \affiliation{\RICE}
\author{G.~Mutchler}
     \altaffiliation[]{\DECEASED}
     \affiliation{\RICE} 

\author {K.~Park} 
     \affiliation{\JLAB}
     \affiliation{\SCAROLINA}
\author {E.S.~Smith} 
     \affiliation{\JLAB}
% ...

\collaboration{The CLAS Collaboration}
\noaffiliation

%
%
\date{\today}
%
%
\begin{abstract}
We searched for the $\Phi^{--}(1860)$  pentaquark in the photoproduction process off the deuteron in the $\Xi^{-} \pi^{-}$  
decay channel using CLAS. The invariant mass spectrum of the $\Xi^{-} \pi^{-}$ system 
does not indicate any statistically significant enhancement near the reported mass $M=1.860~\textmd{GeV}$. The statistical 
analysis of the sideband-subtracted mass spectrum yields a $90\%$ confidence level upper limit of $0.7~\textmd{nb}$ for 
the photoproduction cross section of $\Phi^{--}(1860)$ with a consecutive decay into $\Xi^{-} \pi^{-}$  
in the photon energy range $4.5~\textmd{GeV}<E_{\gamma}<5.5~\textmd{GeV}$. 
\end{abstract}

%
%
%
%
\maketitle
%
%
%
%
\section{Introduction  \label{sec:intro}}
Narrow bound states of four quarks and one anti-quark have been the focus of intense
searches since the report by the LEPS collaboration of a positively charged baryon called the $\Theta^+$, 
with $S=+1$ and  a mass of 1.54 GeV \cite{Nakano:2003qx} . This ``exotic" combination of  quantum numbers cannot
be accommodated within the simple quark model, which assumes that all baryons are built out of three quarks.
Exotic states of this kind have been predicted within the Chiral Soliton Model \cite{Diakonov97}
as part of a spin 1/2 anti-decuplet of baryons. The anti-decuplet of ``pentaquarks"
contains three explicitly exotic states,  whose quantum numbers require a minimal 
quark content of four quarks and one anti-quark. Reference~\cite{Hicks:2005gp} describes the 
experimental situation for $\Theta^{+}(1540)$ searches, and a concise summary of the current state  of pentaquarks
can also be found in the Particle Data Group (PDG) review \cite{Nakamura:2010zzi}.

The two other exotic states of the anti-decuplet have charge $Q = -2$ (quark content of $ddss\overline{u}$) and 
$Q = +1$ (quark content of $uuss\overline{d}$). Their strangeness is $S = -2$,
but they have isospin 3/2, in contrast to normal cascade states with isospin 1/2. The Particle Data Group 
\cite{Nakamura:2010zzi} has assigned the name of $\Phi$(1860) to the four states in the strangeness $S = -2$ 
sector of the anti-decuplet.
The NA49 collaboration has reported evidence for the strangeness $S=-2$ pentaquark  $\Phi^{--}$ 
and the $\Phi^{0}$ at a mass of $1.862$~GeV \cite{Alt04}.  This measurement was conducted in $p+p$
collisions at a center-of-mass energy $\sqrt{s} =17.2~\textmd{GeV}$, and the states were reconstructed 
from their decays into the ground state cascades, $\Phi^{--}\rightarrow \Xi^{-} \pi^-$ 
and $\Phi^{0} \rightarrow \Xi^{-} \pi^+$.
We also note that many experiments \cite{Adamovich:2004yk,Airapetian:2004mi,Chekanov:2005at,Ageev:2005ud,Knopfle:2004tu,
Aubert:2005qi, Stenson:2004yz, Link:2007vy, Abulencia:2006zs,Schael:2004nm,Achard:2006be,Aktas:2007dd,Abdallah:2007bv,
Aleev:2007zz}, some of which represent a much larger statistical sample, have not been able to confirm the NA49  
observation \cite{Alt04}. 

Guidance for where and how to search for cascade pentaquarks is very sparse. 
The mass scale for the $\Theta^+$ can be estimated to be about the mass of the nucleon (0.94 GeV) plus the mass of the
kaon (0.5 GeV). The mass of the cascade pentaquark contains an additional strange quark, 
 which naively would lead to 1.89~GeV, assuming that the strange and anti-strange quarks have the same mass of 
about 0.45 GeV. The quark model predictions vary depending on the amount of mixing between the anti-decuplet and 
octet members, as well as the estimated size of the color-spin hyperfine interaction between quarks. 
We also note that some of the models \cite{Diakonov:2003jj,Jaffe:2003sg,Karliner:2003dt} have used the experimental reports of either
the $\Theta^+(1540)$ or $\Phi(1860)$, or both, to set the scale, thus they are not entirely unbiased.
Table \ref{tab:xi_masses} summarizes representative model predictions 
for the masses of the cascade pentaquark that range from $1.75$ to $2.07~\textmd{GeV}$. 

\begin{table}[t]
\begin{center}
 \begin{tabular}{|l|c|} \hline
Source &  Mass (GeV)  \\ 	\hline
Chiral-Soliton Model \cite{Diakonov97,Diakonov:2003jj}  & 2.07 [1.86] \\ 
Chiral-Soliton Model \cite{Ellis:2004uz}  & 1.79-1.97 \\ 
Diquark Model \cite{Jaffe:2003sg,Jaffe:2003ci}  & 1.75 \\ 
Diquark-Triquark Model \cite{Majee:2007gi,Karliner:2003dt}  & 1.783 \\ 
Experiment NA49 \cite{Alt04}  & $1.862 \pm 0.002$ \\ 
\hline
 \end{tabular} 
\end{center}
\caption{Selected representative model expectations for the mass of the cascade pentaquark. Note that the first
entry gives the initial prediction of the Chiral-Soliton model of 2.07 GeV, followed in square brackets
with the adjusted model to the experimental value of NA49.}
\label{tab:xi_masses}
\end{table} 

This experiment was mounted at Jefferson
Lab to search for the first time for the  $\Phi^{--}$ exotic pentaquark state in real photoproduction off  a neutron  target 
with the subsequent decay into a final state containing three pions and one proton.
Early experimental reports on the $\Theta^+$  suggested that photon beams were a rich source of pentaquarks, and
one calculation predicted that the production of the $\Phi^{--}$  off the neutron was an order of magnitude larger than off the proton \cite{Liu04,Ko:2004jf}.
Liu and collaborators computed  the photoproduction cross section  
$\sigma(\gamma n\rightarrow K^+K^+ \Phi^{--})$ and  $\sigma(\gamma p\rightarrow K^0K^0 \Phi^{+})$,
which correspond to similar reaction channels for the neutron and proton.
At $E_{\gamma} = 5~\textmd{GeV}$ the estimated  cross section $\sigma(\gamma n\rightarrow K^+K^+ \Phi^{--})$,
assuming positive parity for the exotic states, is between 0.4 and 1.5 nb,
depending on the value of the $g_{K^*N\Xi}$ coupling \cite{Liu04}. 
There exists a large range of predictions for the decay widths and branching ratios of these exotic
states \cite{Carlson:2003wc,Oh:2003fs,Diakonov:2003jj}, but the dominant 
decay mode is expected to lead to the ground state cascade $\Phi^{--}\rightarrow \pi^- \Xi^-$. 
In addition, the bias is that the states are very narrow and therefore long-lived, which is particularly
interesting, because they are above the free particle decay thresholds. Therefore, our search was  targeted to
identify states with intrinsic widths that are smaller than the experimental resolution.

\section{Experiment  \label{sec:experiment}}

The purpose of this experiment is to search specifically for the $\Phi^{--}(1860)$ state of the spin-$\frac{1}{2}$ anti-decuplet with 
the CLAS detector \cite{Mecking:2003zu} in Hall B at Jefferson Lab in a photoproduction experiment. The acceptance 
and resolution of CLAS  is better for charged than for neutral particles. The most promising topology 
for our experiment results 
from the decay sequence
\begin{eqnarray}
 \Phi^{--}\rightarrow \pi^- \Xi^- \rightarrow \pi^- (\pi^- \Lambda) \rightarrow \pi^- \pi^-(\pi^- p)~. 
 \label{eq:dec_seq}
\end{eqnarray}
The bremsstrahlung photon beam produced by a $5.77~\textmd{GeV}$ electron beam interacts with the deuteron target, producing a 
large variety of final states.  The outgoing particles are detected and reconstructed in the CLAS detector. The  
energy and the interaction time of the initial photon is determined by registering the electron in the Hall B photon tagging facility 
\cite{Sober:2000we}. 
 
The analysis strategy is to directly reconstruct the decay sequence~(\ref{eq:dec_seq}) from the final state particles
detected in CLAS. 
First, we  identify the $\Lambda(1116)$ using the proton and a $\pi^{-}$. Then we  search for the $\Xi^{-}(1321)$ 
by combining the $\Lambda(1116)$ with another negative pion. Finally, we analyze the invariant mass of  the $\Xi^{-}(1321)\pi^{-}$ 
composite system to search for the  $\Phi^{--}$ pentaquark state.
 
This CLAS experiment collected data during $40$ calendar days at the end of 2004 and 
the beginning of 2005, amounting to approximately $25~\textmd{pb}^{-1}$ integrated luminosity in the tagged photon energy 
range $4.5~\textmd{GeV} \le E_{\gamma} \le 5.5~\textmd{GeV}$. In order 
to achieve an adequate experimental sensitivity in a reasonable amount of time, we operated at an instantaneous  photon flux 
significantly larger than ever used before with CLAS.  The experimental data was carefully analyzed and cross-checked against 
known cross sections  to properly take rate effects into account.

\section{Apparatus   \label{sec:apparat}}

Hall B at Jefferson Lab houses a photon-tagging system \cite{Sober:2000we} to conduct experiments with 
real photons. This facility allows for absolute cross section measurements over a broad energy range of 
the incoming photons. The bremsstrahlung photon beam is produced by the electromagnetic 
radiation of the primary  electron beam in a thin ($\sim 5 \times 10^{-4}$~r.l.) radiator. For this experiment, 
we used the tagged bremsstrahlung beam in Hall B incident on a $40~\textmd{cm}$ long and $4~\textmd{cm}$  
diameter liquid-deuterium target, which was located on the beam axis $50$~cm upstream of the center of the CLAS detector. 

\begin{figure}
\begin{center}
\epsfig{file=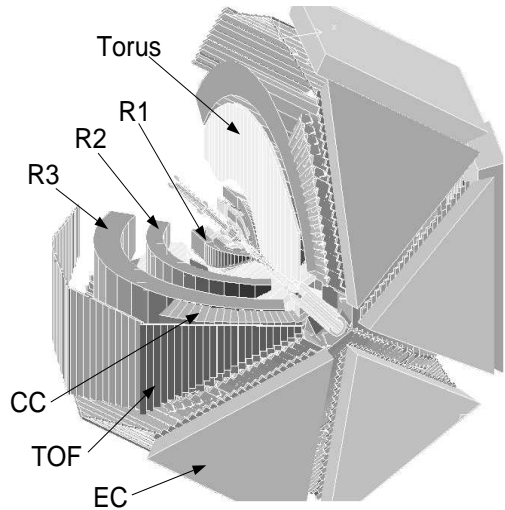, totalheight=7cm, width=7cm, angle=0}
\caption[Three-dimensional view of CLAS]{Three dimensional view of CLAS showing the three regions of Drift Chambers (R1 -R3), 
Cherenkov Counters(CC), the Time-Of-Flight system (TOF) and the Electromagnetic calorimeter (EC) (see text for details). On 
this picture, the photon beam travels from the upper-left corner to the lower-right corner. }
\label{fig:clas}
\end{center}
\end{figure}

CLAS (see Fig.~\ref{fig:clas}) is a nearly $4\pi$ detector that is well-suited to study reactions  
into final states with multiple charged particles. 
The  magnetic field of CLAS \cite{Mecking:2003zu} is provided by six 
superconducting coils, which produce an approximately toroidal field  in the 
azimuthal direction around the beam axis. The regions between the cryostats 
are instrumented with  six identical 
detector packages, also referred to  as  ``sectors''. 
Each sector consists of four Start Counter (ST) paddles 
\cite{Sharabian:2005kq} mainly used for triggering purposes, three regions of 
Drift Chambers (R1, R2, and R3) \cite{Mestayer:2000we}  to determine the trajectories 
of the charged particles, {\v C}erenkov Counters (CC)  \cite{Adams:2001kk} 
for electron identification, Scintillator Counters (SC) \cite{Smith:1999ii} 
for charged particle identification based on  the Time-Of-Flight  
(TOF) method, and Electromagnetic Calorimeters (EC) \cite{Amarian:2001zs} used 
for electron identification and detection of neutral particles. 

The CLAS detector provides a $\frac {\delta p} {p} \sim 0.6\%$ momentum resolution \cite{Mecking:2003zu} 
and  up to $80\%$ of $4 \pi$  solid-angle coverage. 
The efficiency for detection and  reconstruction 
of charged particles in  fiducial regions of CLAS is greater than $95\%$. 
The combined information from the tracking in Drift Chambers and Scintillator Counters allows 
us to reliably separate protons from positive pions for momenta up to $3$~GeV.

\section{Analysis  \label{sec:analys} }

\subsection{Event selection  \label{sec:evt_select}}
One of the main goals of the analysis procedure is to select events corresponding to the reaction
\begin{eqnarray}
 \gamma d \rightarrow \Phi^{--} X,
\end{eqnarray}
where we consider the decay sequence~(\ref{eq:dec_seq}), and kaons in the final state
are not required to be reconstructed.

%
%
\begin{figure}
\begin{center}
\epsfig{file=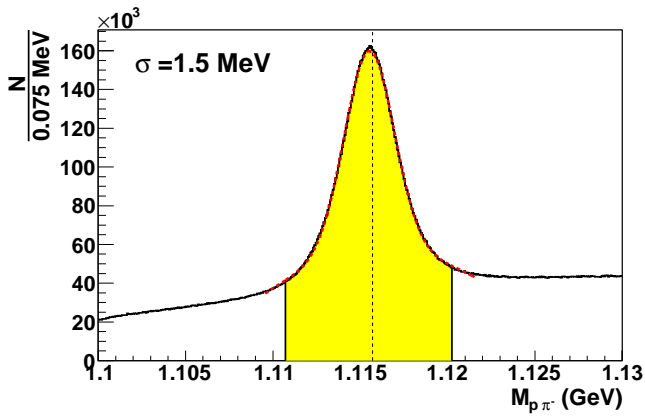, width=8.7cm}
\caption{[Color online] Mass spectrum of the $p\pi^{-}$ system. The yellow shaded area indicates the mass range 
used in this analysis. The red curve is the fit to a Gaussian peak with a polynomial background. 
The dashed vertical line shows the PDG \cite{Nakamura:2010zzi} value for the mass of $\Lambda(1116)$. }
\label{fig:lambda_mass}
\end{center}
\end{figure}

The $\Lambda(1116)$ candidates are identified by considering every pair of positive and negative  
tracks with a hypothesis that these are the proton and the negative pion from  a 
$\Lambda(1116)$ decay using timing information from the scintillation counters  and 
momentum and vertex information from tracking. To select the $\Lambda$-hyperon,  an 
invariant mass cut  $1.1108$~GeV~$<M_{p \pi^{-}}<1.1202$~GeV is applied as shown in Fig.~\ref{fig:lambda_mass}. 
Because the decay products originate from the same point in space,  a 
$5$~cm cut is applied on the Distance-Of-Closest-Approach (DOCA) for the two tracks.  
We define DOCA as the length of the shortest line segment connecting the trajectories of 
these two tracks in the vicinity of the CLAS target. The detector resolution for 
the DOCA between the proton and the $\pi^{-}$ is  $\sim 1.5~\textmd{cm}$. 
This cut reduces the contributions from $p\pi^{-}$  pairs that  
do not come from the  $\Lambda(1116)$ decay.
If there is more than one $\Lambda(1116)$ candidate, we choose the best pair  
based on the combined information from the matching of the invariant mass and the DOCA between the two tracks.

\begin{figure}
\begin{center}
\epsfig{file=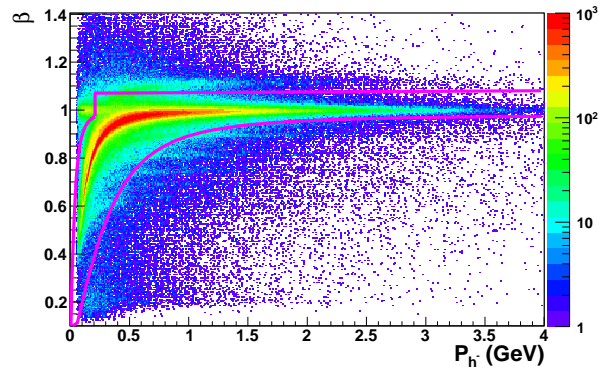, width=8cm}
\caption{[Color online] $\beta$ versus momentum of the negative hadrons. The pions from decays of the $\Lambda$ candidates 
 are not included in this plot. The magenta lines indicate the $\pi^{-}$ identification cuts applied in 
 this analysis.}
\label{fig:pion_id}
\end{center}
\end{figure}

After selecting the best candidate pair for the $\Lambda(1116)$, we proceed with combining it with 
the remaining negative pions in the event, which are identified using time-of-flight and 
tracking information, the event start time determined from the vertex time of the already 
reconstructed $\Lambda(1116)$, and the reference RF-time from the accelerator's injector. 
Figure~\ref{fig:pion_id} illustrates  the negative pion identification 
used in this analysis. The main band corresponds to the negative pions. The magenta lines show 
the cuts applied to select the remaining $\pi^{-}$'s in 
the event. For further analysis, we require that an event contains 
at least two more negative pions in addition to the $\pi^{-}$ in the $\Lambda(1116)$ pair. 
Therefore, one can have multiple combinations of ($\Lambda \pi^{-}$) pairings, and we 
considered all combinations of the $\Lambda$ and each of the remaining pions in the event whose 
track separation in space (DOCA) was sufficiently small.   
This treatment of the $\Lambda \pi^{-}$ pairs may lead to multiple entries in the background, 
but counts the correct pairing in the cascade peak only once. 

We also require that the time reconstructed from the event in CLAS matches 
the time of the interaction determined by the information from the photon tagger. If there is 
at least one photon detected in the tagger that  can provide enough  energy for the 
$d \left( \gamma , \Xi^{-} \pi^{-} \right ) K^{+} K^{+}p$ reaction for the measured kinematics, and it is 
registered within $\pm 3.2~\textmd{ns}$ of the interaction time from CLAS,  then this event 
is kept in the data sample. 
Table~\ref{tab:cons_cuts} summarizes the event selection cuts used in this analysis.  

%
%

\begin{figure}
\begin{center}
\epsfig{file=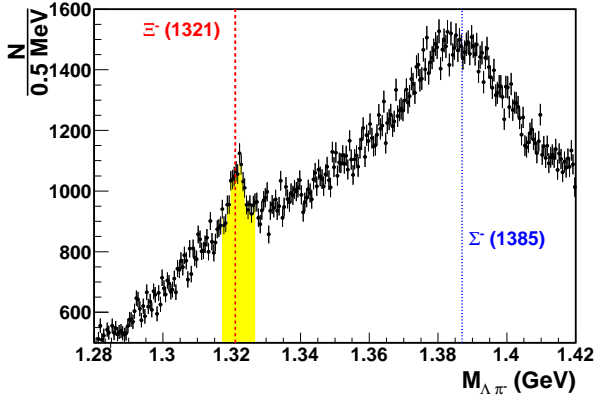, width=8cm}
\caption{[Color online] Invariant mass of $\Lambda \pi^{-}$ pairs. The red dashed and blue dotted lines mark the positions of 
the nominal mass for $\Xi^{-}(1321)$ and $\Sigma^{-}(1385)$, respectively \cite{Nakamura:2010zzi}. The shaded 
area shows the mass range used in this analysis. }
\label{fig:xi_mass}
\end{center}
\end{figure}

\begin{figure}
\begin{center}
\epsfig{file=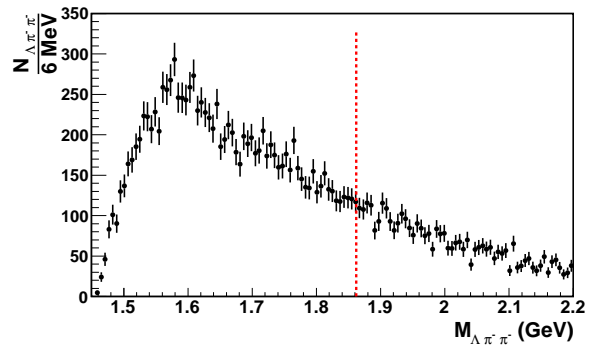, width=8cm}
\caption{[Color online] Invariant mass of $\Lambda \pi^{-} \pi^{-}$ after applying the cascade mass cut. The dashed line marks 
the position of the reported mass of $\Phi^{--}(1860)$.}
\label{fig:lbdpipi_mass}
\end{center}
\end{figure}

\begin{figure}
\begin{center}
\epsfig{file=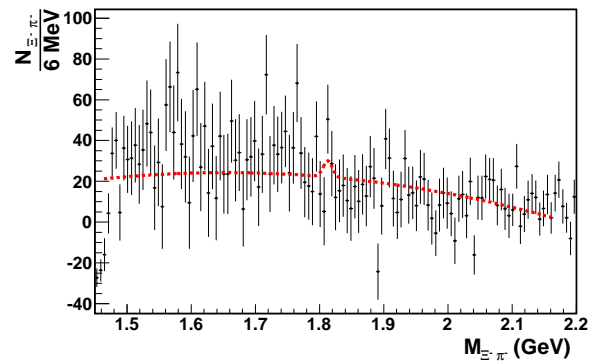, width=8.0cm}
\caption{[Color online] Number of  $\Xi^{-} \pi^{-}$ events per $6~\textmd{MeV}$ mass bin. The error 
bars indicate the statistical uncertainties. The  dashed curve shows the fit to a Gaussian peak 
above a polynomial background. The center of the Gaussian in this plot is fixed at the center of the bin 
at $1.813~\textmd{GeV}$. }
\label{fig:mass_xi_pi}
\end{center}
\end{figure}

Figure~\ref{fig:xi_mass} shows the 
invariant mass spectrum of the $\Lambda \pi^{-}$ pairs after the cuts from Table~\ref{tab:cons_cuts}, with the   
exception of the cascade mass cut and the DOCA cut in the last two rows. 
The red dashed and blue dotted lines show the positions of 
the nominal mass for the $\Xi^{-}(1321)$ and $\Sigma^{-}(1385)$, respectively \cite{Nakamura:2010zzi}. We 
apply the $1.3175~\textmd{GeV} \le M_{\Lambda \pi^{-}} \le 1.3265~\textmd{GeV}$ cut illustrated by the shaded area 
to select events with cascade hyperons.  
Figure~\ref{fig:lbdpipi_mass} shows the invariant mass of
$\Lambda \pi^{-} \pi^{-}$ after the cascade mass cut and the cut on the DOCA between the $\Xi^{-}$ candidate and 
the negative pion described in Table~\ref{tab:cons_cuts}. There is no statistically significant structure 
near the reported mass of the $\Phi^{--}(1860)$. We use the sideband subtraction method to account for the background 
contribution in the $\Lambda \pi^{-} \pi^{-}$ mass spectrum coming from events under the cascade peak in 
Fig.~\ref{fig:xi_mass}. 
The $\Lambda \pi^{-} \pi^{-}$ spectra from the mass ranges 
$1.300~\textmd{GeV} \le M_{\Lambda \pi^{-}} \le 1.310~\textmd{GeV}$ and 
$1.335~\textmd{GeV} \le M_{\Lambda \pi^{-}} \le 1.345~\textmd{GeV}$
normalized to the number of background events under the cascade peak are subtracted from the mass spectrum in 
Fig.~\ref{fig:lbdpipi_mass} to obtain the sideband-subtracted $\Xi^{-} \pi^{-}$ mass distribution shown in 
Fig.~\ref{fig:mass_xi_pi}.

\begin{table}
\begin{center}
 \begin{tabular}{|l|c|c|} \hline
DOCA($\Lambda$)       &  DOCA($\Lambda$) $\le$ 5.0 cm   \\ 
			\hline
Mass($\Lambda$)       &  $1.1108~\textmd{GeV} \le M_{\Lambda} \le 1.1202~\textmd{GeV}$   \\ 
			\hline
DOCA($\Xi$)           &  DOCA($\Xi$) $\le$ 4.5 cm   \\ 
			\hline
$\beta$ vs $p$         &  $p$-dependent cut shown in Fig.~\ref{fig:pion_id}   \\ 
			\hline
Tagger time           &  $-3.2~\textmd{ns} \le T_{\gamma} - T_{vtx} \le  +3.2~\textmd{ns}$  \\ 
			\hline
Missing Mass         &  $MM_{\Xi^{-} \pi^{-} } > 2 M_{K} + M_{p}$  \\ 
			\hline 
Mass($\Xi$)           &  $1.3175~\textmd{GeV} \le M_{\Lambda \pi^{-}} \le 1.3265~\textmd{GeV}$   \\ 
			\hline 
DOCA($\Xi \pi$)       &  DOCA($\Xi \pi$) $\le$ 4.5 cm   \\ 
			\hline
 \end{tabular} 
\end{center}
\caption{Summary of the event selection.}
\label{tab:cons_cuts}
\end{table}

%
%
%

%
%
%

\subsection{Detector simulation  \label{sec:simul}}
In order to relate the experimental yields to cross sections,  acceptance correction 
factors were calculated using the Monte-Carlo method. The GEANT-based detector simulation 
package  incorporates the survey geometry of CLAS, realistic response of drift chamber and scintillation 
counters, as  well as documented inefficiencies due to dead wires and malfunctioning photomultiplier tubes. 
Because CLAS is a complex detector covering almost a $4\pi$  solid angle, it is virtually impossible 
to separate the efficiency 
calculations from the geometrical acceptance calculations. In this paper, the term acceptance 
correction refers to a combined correction factor due to the geometry of the detector and 
the inefficiencies of the detection and reconstruction. It is defined as the ratio of the 
number of reconstructed Monte-Carlo events to the number of  simulated  events in each given kinematic bin.

\begin{figure}
\begin{center}
\epsfig{file=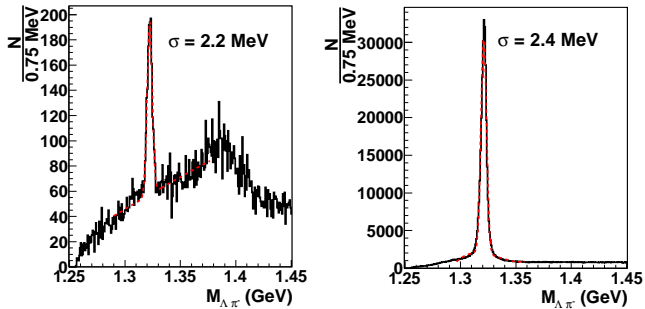, width=8.5cm}
\caption{[Color online] Mass spectrum of $\Lambda \pi^{-}$  from data (a) using restrictive cuts (see text for 
details), and $\Lambda \pi^{-}$ mass from the GEANT-based simulations (b). }
\label{fig:mass_data_sim}
\end{center}
\end{figure}

The event sample used in the acceptance calculation was generated using a phase-space generator with an 
event configuration in the first row in Table~\ref{tab:accp_evts} without any physics background. 
Figure~\ref{fig:mass_data_sim} shows the invariant mass of the $\Lambda \pi^{-}$ system from the data (a) and 
from the simulations (b). The simulated event sample does not contain any background because we are 
interested only in determining the acceptance and efficiency for the events which contain cascade hyperons. 
The acceptance and efficiency corrections calculated for events containing $\Xi^{-} \pi^{-}$ are applied to the 
sideband-subtracted spectrum. 
Event selection criteria for the  plots in Fig.~\ref{fig:mass_data_sim} are stricter than 
the nominal cuts in this analysis to enhance the signal-to-background ratio for the ground state cascade peak 
for a visual comparison of the data with the simulations. 

%

\subsection{Model dependence  \label{sec:model_dep}}
Because we do not know how the cross section of the $\Phi^{--}(1860)$  photoproduction depends on kinematics, 
and since we integrate over all of the kinematic variables,  our estimate of the CLAS acceptance 
depends on the choice of the distribution of the events over the accessible  phase space.  
In order to estimate the uncertainty of the acceptance due to the model dependence, we studied 
four event configurations; the relative acceptances are given in Table~\ref{tab:accp_evts}. 
\begin{table}
\begin{center}
 \begin{tabular}{|c|l|c|} \hline 
Row & Production            & Relative   \\ 
\#  & Model                 & Acceptance \\ \hline
1   & $\gamma d \rightarrow K^{+} K^{+} \Xi^{-} \pi^{-} p_{s}$  & ~~1.00 \\ \hline
2   & $\gamma d \rightarrow K^{+} \Sigma^{-}(2650) p_{s}$ & \\
    &  \hspace{0.35cm} $ \rightarrow K^{+} K^{+} \Phi^{--} p_{s} \rightarrow   K^{+} K^{+} \Xi^{-} \pi^{-} p_{s}$  & +1.24 \\  \hline
3   & $\gamma d \rightarrow K^{+}_{f} \Sigma^{-}(2650) p_{s} $ &  \\
    &   \hspace{0.35cm} $ \rightarrow K^{+}_{f} K^{+} \Phi^{--} p_{s} \rightarrow   K^{+}_{f} K^{+} \Xi^{-} \pi^{-} p_{s}$  & +1.47  \\  \hline
4   & $\gamma d \rightarrow K^{+} K^{+} \Phi^{--} p_{s} \rightarrow  K^{+} K^{+} \Xi^{-} \pi^{-} p_{s}$ & +1.07  \\ \hline
 \end{tabular} 
\end{center}
\caption{Event configurations used in the Monte-Carlo generator for the model dependence studies.
The elementary production is assumed to be from a neutron in a deuterium target and the proton, $p_s$, is treated as a
spectator in the reaction.}
\label{tab:accp_evts}
\end{table} 
The events were generated using a software package that includes essentially no specific dynamics and 
mainly simulates events according to phase-space probabilities.
The reaction in row 1 of  Table~\ref{tab:accp_evts} is a four-body 
phase-space uniform distribution complemented with a 
spectator proton with  Fermi momentum smearing according to Ref.~\cite{Gibbs:2004ji}. For the process in row 2, a hypothetical
$\Sigma^{-}(2650)$ was implemented in the event generator, which decays into $K^{+} \Phi^{--}(1860)$ with a total 
width of $35$~MeV. The $\Phi^{--}(1860)$ is simulated as a particle with an infinitely narrow width at  mass 
$M_{\Phi}=1.862~\textmd{GeV}$ that decays only through the $\Xi^{-} \pi^{-}$ channel. The reactions in rows 2 and 4
are simulated according to two- and three-body phase space, respectively, with a Fermi-smeared spectator proton momentum 
spectrum according to Ref.~\cite{Gibbs:2004ji}.  The process in 
row 3 of Table~\ref{tab:accp_evts} is simulated according to two-body phase-space but with an additional 
exponential $t$-dependence for the $K^{+}$  with a $t$-slope parameter $b=2.6~\textmd{GeV}^{-2}$.  
The flat phase space provides an estimate of the acceptance for  
$s$-channel processes, while including a steeper $t$-dependence  allows us to consider 
the processes going through the $t$-channel exchanges. 
For the processes in rows  2 to 4, the acceptance 
is estimated at a fixed mass $M_{\Xi^{-} \pi^{-}}=1.862~\textmd{GeV}$, since the process explicitly 
includes the $\Phi^{--}(1860)$ decay. 
Using this table, we calculate the root-mean-square (RMS) of the differences in the acceptance at $M_{\Xi^{-} \pi^{-}}=1.862~\textmd{GeV}$, 
and we assign a relative uncertainty of $\frac{\sigma^{sys}_{A}} {A} \sim 21\%$ 
due to model dependence. 
This is  the largest source of uncertainty in the determination of the cross sections and 
its upper limits. 

\begin{figure}
\begin{center}
\epsfig{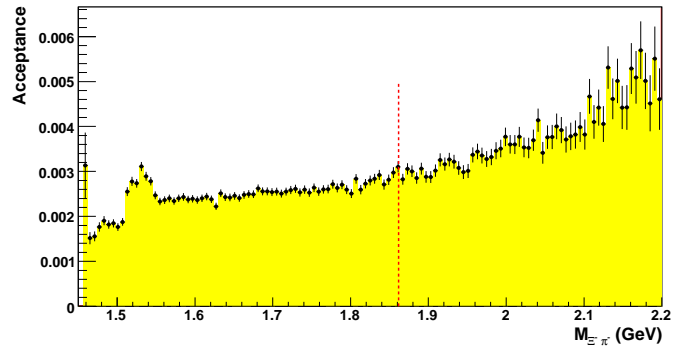}
\caption{[Color online] Acceptance of CLAS versus the invariant mass of the $\Xi^{-} \pi^{-}$ system, determined from 
the $\gamma d \rightarrow K^{+} K^{+} \Xi^{-} \pi^{-} p_{s}$  phase space simulation, using a GEANT-based 
detector simulation package GSIM. The red line indicates the position of the peak seen by the NA49 
collaboration. }
\label{fig:accp_mass}
\end{center}
\end{figure}

The acceptance of CLAS versus the invariant mass of the $\Xi^{-} \pi^{-}$ system calculated using the process in row 1 
of Table~\ref{tab:accp_evts} is shown in  Fig.~\ref{fig:accp_mass}. 
At the expected $\Phi^{--}$ mass $M=1.862~\textmd{GeV}$,  the acceptance does not exhibit any special features and 
is $\sim 0.3\%$. The artificial enhancement near mass $M=1.53~\textmd{GeV}$ is due to pion combinatorics and  
the mass cut to select $\Xi^{-}(1321)$ events. 

%
%

\subsection{Trigger conditions and normalization \label{sec:trig_inef} }
The process of interest for this experiment is $\gamma d \rightarrow p \pi^{-} \pi^{-} \pi^{-} X$, 
where three charged pions and a proton are detected in CLAS. To be able to run at higher 
photon-deuteron luminosities, we used a highly selective trigger, so that the data acquisition system could
cope with the data rate. After extensive studies, we decided to require at least  
three charged particles in three different sectors of CLAS to be detected in the Start Counter and TOF system in the main trigger. 
But we also took data with a two-sector trigger to quantify trigger inefficiencies, which
was prescaled by a factor of 5 to 20, depending on the running conditions.
The trigger required a time coincidence of tracks in CLAS with a signal from the photon tagger, signaling
the production of a photon in the energy range of 4.5 to 5.5 GeV. 

The primary trigger condition suffered from inefficiencies at high luminosity. These inefficiencies were
determined empirically by studying the luminosity dependence of $\Lambda$ production and also by comparing
these yields with those measured with the two-sector trigger, corrected for the prescale factor.
The study used events containing an identified $\Lambda$-hyperon, decaying to a proton and a pion, and two additional reconstructed 
tracks, similar to our sample of signal events.  
The average inefficiency of the three-track trigger varied linearly with the electron beam current up to 
$35\%$, with an average of $25\%$ at the nominal current of $30~\textmd{nA}$. 
The yields normalized by this factor correspond to those 
obtained from the two-track trigger, which did not exhibit any dependence on luminosity.

The overall normalization of the experiment was checked using the $d(\gamma, \pi^{-} \Delta^{++})n$ reaction 
with the detection of the $\pi^{+}\pi^{-} p$ final state. 
The cross section of this process is expected to be mostly dominated by  photoproduction off a quasi-free proton 
$\gamma p \rightarrow \pi^{-} \Delta^{++}$. 
Although final state interactions (FSI) contribute to this process, we do not expect their impact to be 
significant within the precision required for this purpose. 
Using the same analysis and assumptions, the measured cross sections for $d(\gamma, \pi^{-} \Delta^{++})n$ obtained from the 
current data was compared to the $p(\gamma, \pi^{-} \Delta^{++})$ cross section from a different CLAS run period where these 
trigger inefficiencies were not present. 
The cross sections from the two CLAS data sets for the photon energy 
range of $4.5~\textmd{GeV} < E_{\gamma} < 5.0~\textmd{GeV}$ differed by  $9 - 14\%$. 
The agreement of our measurements with the published data on $\gamma p \rightarrow \pi^{-} \Delta^{++}$ from SAPHIR at
$E_\gamma$=2.5 GeV \cite{Wu:2005wf} is better than $10\%$, which indeed indicates that FSI contributions are negligible. 
A more detailed analysis of the $d(\gamma, \pi^{+} \pi^{-} p)n$ reaction and its cross section is under 
way \cite{Graham:xxxxxx}. Based on these comparisons, we have assigned the normalization uncertainty  of $\pm15\%$ for the 
presented  data.

\subsection{Systematic uncertainties \label{sec:sys_err} }
The final invariant mass spectrum of $\Xi^{-} \pi^{-}$ was studied for various values 
of the selection parameters for the $\Lambda$, $\Xi$, and $\pi^{-}$. These 
variations did not result in any qualitative change of the invariant mass distributions
of $\Lambda \pi^{-} \pi^{-}$ or $\Xi^{-} \pi^{-}$. 
We checked the sensitivity of the results with respect to the following parameters: 
selection parameters for the $\Lambda$-candidates, cuts on time matching between the event 
in CLAS and hit in the tagger counter,  DOCA cuts on the $\Lambda$ and the $\Xi^{-}$ candidates, 
 detached vertex cuts for the reconstructed $\Lambda$, and particle identification cuts on $\beta$ 
versus $p$.  The choice of these analysis parameters did not 
affect the final result for the upper limit of the cross sections, and  we therefore 
considered their uncertainties to be negligible. 

The uncertainty  in the overall normalization of the experiment is driven by the relatively large
trigger efficiency corrections that  are required, as described previously.
We assign an uncertainty of $\pm 15\%$ to the absolute normalization, which is estimated by comparing our 
determination of $\Delta^{++}$ production to other measurements, as discussed in the previous section. 

The dominant contribution to the photoproduction cross section systematic uncertainty comes
from the model dependence of the estimated acceptance, and we take the RMS of the  model calculations 
to obtain a $\pm 21\%$ relative uncertainty. 

Table~\ref{tab:sys_errors} summarizes the dominant contributions to the systematic uncertainties. 
We used the sum in quadrature  of $26 \%$ to determine the upper limits for the experimental cross 
sections described in the next section.

\begin{table}[]
\begin{center}
 \begin{tabular}{|l|c|} \hline
Source of Uncertainty &  Uncertainty  ($\sigma$)      \\  \hline  
Model dependence of acceptance    &  $21\%$           \\          
Flux and trigger efficiency       &  $15\%$           \\  \hline           
Total in quadrature               &  $26\%$           \\  \hline
 \end{tabular} 
\end{center}
\caption{Sources and the values of the relative systematic uncertainties. The overall systematic 
uncertainty calculated as the square root of the quadrature sum is $\pm 26\%$.}
\label{tab:sys_errors}
\end{table}

\section{Upper limits  }
\label{sec:upper_limits}

In order to determine the upper limits for the cross section for a possible peak, we scanned each $6~\textmd{MeV}$-wide
bin in the sideband-subtracted mass spectrum in Fig.~\ref{fig:mass_xi_pi}, considering the center 
of each bin as the mean value of a Gaussian distribution with a fixed width of $\sigma_{G}=7~\textmd{MeV}$,
which represents our experimental resolution for a potential $\Phi^{--}(1860)$, as 
estimated by the detector simulation. 
Then the points in the neighborhood of the bin are fitted to a Gaussian peak plus a polynomial. 
The purpose of this $\chi^{2}$-fit is to provide us with an estimate of the background 
under the possible peak and its uncertainty. The total number of signal events is calculated  
as the excess of the observed events over the fitted background, both integrated within a window of $20~\textmd{MeV}$ 
around the center of each bin. The red dashed curve in Fig.~\ref{fig:mass_xi_pi} shows a particular 
example of the fit for a  mass bin centered at $M_{\Xi^{-} \pi^{-}}=1.813~\textmd{GeV}$. 

In order to obtain an upper limit on the photoproduction cross section, we developed 
a procedure based on the method described in Ref.~\cite{Rolke:2004mj}. This procedure allows us to factor in the 
uncertainties in the background 
extraction and the acceptance correction into the determination of the upper limits at a given confidence level (CL)
using the sideband-subtracted method. 
We also performed a cross check of our method with an  approach for estimating the upper limits by Smith~\cite{Smith:2008fu} 
based on the construction prescription of the Feldman-Cousins method \cite{Feldman:1997qc}, which 
properly takes into account the systematic uncertainties when constructing the confidence belts. 
For a comparison, we assumed a mass-independent acceptance of $0.4\%$ and a mass-independent 
relative acceptance uncertainty of $30\%$; the agreement between the two methods is very good.
A more detailed description of our method to determine the upper limits can be found in Appendix A.

\begin{figure}
\begin{center}
\epsfig{file=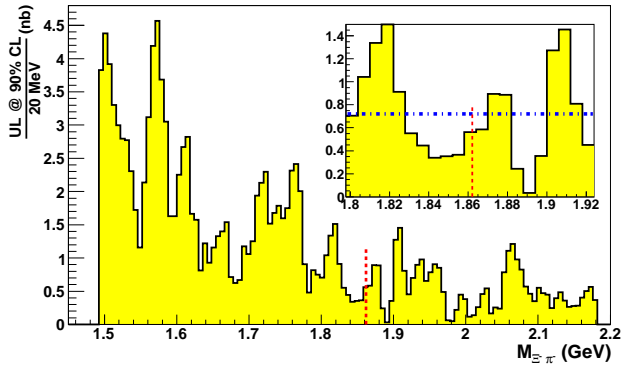, width=9cm, angle=0}
\caption{[Color online] Upper limits for the photoproduction cross section with subsequent 
decay $\Phi^{--} \rightarrow \Xi^{-} \pi^{-}$ in $20$~MeV mass windows at $90\%$ CL versus the
invariant mass of $\Xi^{-} \pi^{-}$. The red dashed vertical line marks the position of the enhancement 
reported by  NA49 . }
\label{fig:ul_cs}
\end{center}
\end{figure}

The upper limit of the photoproduction cross section at $90\%$ confidence level for the process 
$\gamma d \rightarrow \Phi^{--} X \rightarrow \Xi^{-} \pi^{-} X$ versus invariant mass of the 
$\Xi^{-} \pi^{-}$ system is shown in Fig.~\ref{fig:ul_cs}. In the mass range near 
$M=1.862~\textmd{GeV}$, where the NA49 collaboration observed an enhancement, we obtain a $90\%$ CL upper 
limit of $\sim 0.7~\textmd{nb}$. 

%
%
%
%

%
\section{Summary \label{sec:summary}}
In conclusion,  we have conducted for the first time a search for the $\Phi(1860)^{--}$ pentaquark state in 
real photoproduction within the incident photon energy range $4.5~\textmd{GeV} \le E_{\gamma} \le  5.5~\textmd{GeV}$. 
We do not observe any statistically significant enhancement near invariant mass  $M=1.862~\textmd{GeV}$. 
The upper limit at $90\%$ confidence for the photoproduction cross section of the reaction $\gamma d \rightarrow \Phi^{--}X$ 
multiplied by the branching ratio for $\Phi^{--}\rightarrow \Xi^{-} \pi^{-}$ is
determined as  a function of  the invariant mass of  $\Xi^{-} \pi^{-}$, using a method similar 
to the one described in Ref.~\cite{Rolke:2004mj}.
The upper limit for the cross sections for $\Phi^{--}$ photoproduction with subsequent 
decay $\Phi^{--} \rightarrow \Xi^{-} \pi^{-}$ for $20$~MeV mass windows 
is less than $3~\textmd{nb}$ in the $\Xi^{-} \pi^{-}$ mass range between $1.6~\textmd{GeV}$ and $1.9~\textmd{GeV}$. 
The upper limit is less  than $1.5~\textmd{nb}$ for the masses from  $1.9~\textmd{GeV}$ to $2.2~\textmd{GeV}$. 
The upper limit for the cross section averaged within a narrow mass range of $1.80$~GeV$~< M_{\Xi \pi } < 1.92$~GeV is 
$\sim 0.7~\textmd{nb}$. This is approximately a factor of three improvement over the previously estimated upper limit in 
small-angle electroproduction by the HERMES collaboration of $\sim 2~\textmd{nb}$ \cite{Airapetian:2004mi}.

\begin{acknowledgements}
We would like to thank the staff of the Accelerator and Physics 
Divisions at the Jefferson Laboratory for their outstanding efforts 
to provide us with the high quality  beam and the facilities for  
data analysis.
This work was supported by the U.S. Department of Energy and the  
National Science Foundation, the French Commissariat \`{a} l'Energie 
Atomique, the Italian Istituto Nazionale di Fisica Nucleare, 
the National Research Foundation of Korea, Chilean CONICYT,
and United Kingdom's Science and Technology Facilities Council. 
The Southeastern Universities 
Research Association (SURA) operated the Thomas Jefferson National 
Accelerator Facility for the United States Department of Energy under 
Contract No. DE-AC05-84ER40150.  The U.S. Government retains a non-exclusive, 
paid-up, irrevocable, world-wide license to publish or reproduce this manuscript 
for U.S. Government purposes. 
\end{acknowledgements}

\bibliographystyle{prsty}
\bibliography{references}

\appendix

\section{Appendix: Determination of upper limits}
\label{apx:upper_limits}

The number of events in the $20~\textmd{MeV}$ mass windows from the sideband-subtracted spectrum in Fig.~\ref{fig:mass_xi_pi}   
are distributed according to a Gaussian distribution with a width determined by the statistical uncertainty obtained 
during the sideband subtraction. 
Therefore, in each mass window we model the excess of the events above the background, or signal events, according to 
a Gaussian distribution, with a mean $\mu \ge 0$ limited by the condition that 
the cross section cannot be negative even if the number of observed events is less than the expected background. 

Systematic uncertainties are included into the calculation by assuming that the measured acceptance and the number of 
background events are random variables distributed according to the normal distribution:
\begin{eqnarray}
P(x, b_{m}, e_{m} ~\vert~ \mu, b, e ) & = &  \frac{1}{\sqrt{2\pi}\sigma_{x}} 
	e^{-\frac{(e \mu+b - x)^2}{2\sigma_{x}^{2}}}   \\ \nonumber 
        & \times & \frac{1}{\sqrt{2\pi}\sigma_{b}}  e^{-\frac{(b_{m} - b)^2}{2\sigma_{b}^{2}}}  \\ \nonumber 
        & \times & \frac{1}{\sqrt{2\pi}\sigma_{e}}  e^{-\frac{(e_{m} - e)^2}{2\sigma_{e}^{2}}} ~, 
\label{eq:rolke_likel}
\end{eqnarray}
where $\mu$  is the expectation of the number of signal events, $e$ is the acceptance factor for the signal, $b$ 
is the expectation value of the number of background events in the 20 MeV mass window, and $x$ is the observed number of events in 
the same  window for each fit. 
The experimental statistical uncertainty is denoted by $\sigma_{x}$. The estimated number 
of background events $b_{m}$ is determined from the 
polynomial fits to the background, as shown in Fig.~\ref{fig:mass_xi_pi}, and its uncertainty, $\sigma_{b}$, is taken from errors returned 
from the fit. The estimated signal acceptance $e_{m}$ is determined by Monte Carlo and shown in Fig.~\ref{fig:accp_mass}. For 
the uncertainty in the value of $e_{m}$, $\sigma_{e}$, we use the systematic uncertainty of $26\%$ (Table~\ref{tab:sys_errors}). 
$P(x, b_{m}, e_{m} ~\vert~ \mu, b, e )$ is the probability, under assumption 
of our model, to observe the values $x$, $b_{m}$ and $e_{m}$.
Similar to the Rolke method in Ref.~\cite{Rolke:2004mj} we use a profile likelihood to estimate the 
confidence level (CL). The logarithm of the profile likelihood is defined as the logarithm of the ratio 
\begin{eqnarray}
   \nonumber \cal{L} &=& -2 \ln{\lambda(\mu_{t})}  \\ 
   \nonumber &=& - 2 \ln{ \left( \frac{ \textmd{sup}\{P(x, b_{m}, e_{m} ~\vert~ \mu_{t}, b, e ); b, e\} } 
	{\textmd{sup}\{P(x, b_{m}, e_{m} ~\vert~ \mu, b, e ); \mu, b, e \}} \right) } \\   
	 &=& \frac{ (\mu_{t} e_{m} + b_{m} - x)^{2}} {\sigma_x^2 + \mu_t^2 \sigma_e^2 + \sigma_b^2}  ~, 
\label{eq:prof_like}
\end{eqnarray}
where $\lambda$ is the profile likelihood, and $\mu_{t}$ is the hypothesis value being tested.
The supremum, or the least upper bound, $\textmd{sup}\{P(x, b_{m}, e_{m} ~\vert~ \mu_{t}, b, e ); \mu, b, e\}$ 
in the denominator under the logarithm in Eq.~\ref{eq:prof_like} is taken over all 
values of $(\mu, b, e)$, and is located at $\left( b=b_{m}, e=e_{m}, \mu=\frac{x-b}{e} \right)$. 
The least upper bound $\textmd{sup}\{P(x, b_{m}, e_{m} ~\vert~ \mu, b, e ); b, e \}$ 
in the numerator is taken only over the background and efficiency $b$ and  $e$. In order to determine  
the location of this supremum we find the zero-crossings of both partial derivatives of the likelihood $P$ in 
Eq.~\ref{eq:rolke_likel}: 
$\frac{\partial P}{\partial b}=0$ and $\frac{\partial P}{ \partial e}=0$. In the limiting case when 
the background is known to be zero and the efficiency is $100\%$,  Eq.~\ref{eq:prof_like} simply becomes
\begin{eqnarray}
  \cal{L} &=& \frac{\left( \mu_{t} - x \right)^{2}}{\sigma_{x}^{2}}~.
\label{eq:prof_like_lim}
\end{eqnarray}

The log-likelihood distribution in Eq.~\ref{eq:prof_like} is approximated by the $\chi^{2}$-distribution ($\chi^{2} \approx \cal{L}$) with the 
appropriate number of degrees of freedom. In this case, there is only 
one degree of freedom. To find the values of $\mu_{t}$ corresponding to a certain confidence level, 
one first finds the $\chi^{2}$ value corresponding to that CL for the $\chi^{2}$-distribution with a single degree of 
freedom. The solutions for $\mu_{t}$ are the values where $\cal{L}$ differs from its minimum by that amount of 
$\chi^{2}$. 
The solution that is less than the most likely value of $\mu$ is the lower limit, while   
the larger solution is the upper limit. After substituting $\chi^{2}$ for $\cal{L}$ in Eq.~(\ref{eq:prof_like}) and 
solving the quadratic equation,
the two solutions for $\mu_{t}$ for a given $\chi^{2}$ can be found as follows: 

\begin{eqnarray}
 \mu_{t}    &=&   \frac{ -b_{m}~e_{m} + x e_{m} } { \chi^2 \sigma_e^2  - e_{m}^2~ } +  
	            \left( e_{m}^2 ~\chi^2 \sigma_x^2 + e_{m}^2 \chi^2 \sigma_b^2   \right. \\ \nonumber
	& - &  2 \chi^2 \sigma_e^2 x b_{m} + \chi^2 \sigma_e^2 x^2 - 
              \chi^4 \sigma_e^2 \sigma_x^2 +\chi^2 \sigma_{e}^2 b_{m}^2  \\ \nonumber 
	& - & \left. \chi^{4} \sigma_{e}^{2}\sigma_{b}^{2} \right) ^{\frac{1}{2}} 
	 \times  \left( \chi^{2} \sigma_{e}^{2} - e_{m}^{2} \right)^{-1} ~.
\label{eq:rolke_sols}
\end{eqnarray}

%
%
\begin{figure}[t]
\begin{center}
\epsfig{file=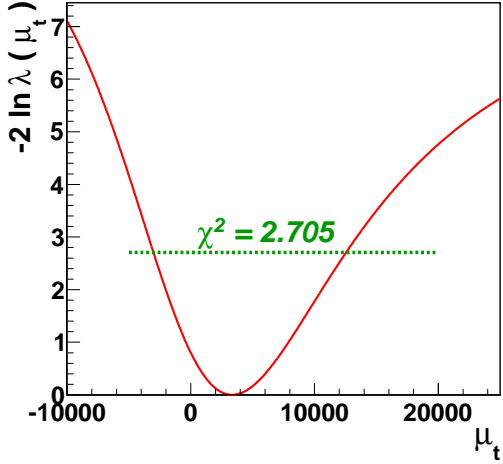, width=7.2cm}
\caption{[Color online] Illustration of determining the upper and lower limits for $CL=90\%$. The $x$-axis 
is the expectation value for the measured quantity $\mu$. The red solid curve is the logarithm of the 
profile likelihood $\cal{L}$  versus $\mu_{t}$, and the green dashed line is $\chi^2=2.705$ 
corresponding to $CL=90\%$.}
\label{fig:rolke_ilus}
\end{center}
\end{figure}
%

Fig.~\ref{fig:rolke_ilus} illustrates how the upper and lower limits at $90\%$ CL are found. 
The red curve is the log-likelihood $\cal{L}$ with a minimum at around $\mu_{t} \sim 3300$,  
which is the most likely value for the illustrated example. The probability of having $90\%$ of the 
trials within a certain range is realized for $\chi^2=2.705$, which is represented by the green line 
in Fig.~\ref{fig:rolke_ilus}.  
The intersection points of these curves give us the upper and lower limits at $90\%$ confidence level. 
In certain cases, for instance when the estimated efficiency is very low, and the uncertainty for 
it is relatively large, no upper or lower limits can be found. This happens when the logarithm of the 
profile likelihood $\cal{L}$ does not behave like a parabola, and therefore is not well approximated by 
the $\chi^{2}$-distribution. 

In cases where the most likely $\mu_{t}$ is negative, we take the lower limit to be $0$. If the most likely 
value $\mu_{t}$ and the upper limit are negative,  we increment the number of observed 
events $x$ by one unit until we get the first positive value. Such an ad hoc adjustment can cause the coverage 
probability, defined as the probability that a true value of the cross section 
for a process is less than the corresponding upper limit obtained by this method, to differ from the desired confidence 
level of $90\%$.  Therefore, one needs to check that the results obtained 
by this procedure indeed provide the desired confidence level. Our Monte-Carlo tests showed that  the coverage 
probability of this method is within $5\%$ of the nominal confidence level.

\end{document}